# CONTRIBUTION OF NANO-SCALE EFFECTS TO THE TOTAL EFFICIENCY OF CONVERTERS OF THERMAL NEUTRONS ON THE BASIS OF GADOLINIUM FOILS

D. A. Abdushukurov[*)], D.V.Bondarenko, Kh.Kh.Muminov[**)], D.Yu.Chistyakov

Physical-Technical Institute of Academy of Sciences of the Republic of Tajikistan, Aini Ave 299/1, Dushanbe 734063, Tajikistan

e-mails: [*)] abdush@tajik.net, [**)] muminov@tascampus.eastera.net

Abstract

We study the influence of nano-scale layers of converters made from natural gadolinium and its 157 isotope into the total efficiency of registration of thermal neutrons. Our estimations show that contribution of low-energy Auger electrons with the runs about nanometers in gadolinium, to the total efficiency of neutron converters in this case is essential and results in growth of the total efficiency of converters. The received results are in good consent to the experimental data.

1. Introduction

Converters of neutron radiation play a determining role at designing detectors of neutron radiation. They determine key parameters of detectors, such as efficiency, the spatial resolution and so on. Among the solid-state converters especially stand out the converters made on the basis of isotopes of $^{6}$Li, $^{10}$B and $^{157}$Gd. These isotopes have abnormal high sections of interaction with neutrons and they are not activated, these features make them rather attractive for their application as converters of neutron radiation. In our researches we were engaged in problems of modeling of characteristics of solid-state converters and we used converters made from natural gadolinium and its 157 isotope in our researches. It is well known, that as a result of radiating capture of thermal neutrons by gadolinium nucleus, electrons of internal conversion and Auger electrons are radiated. Namely these secondary electrons are registered basically by the detectors. Converters made from gadolinium are used in both gas-filled and solid-state detectors. For the first time gadolinium converters in position-sensitive detectors were used in works /1,2/. In the same work the results of modeling of the efficiency of gadolinium converters are discussed. In these calculations only electrons with the energies higher than 29 keV were taken into account. It was related to the reason that as the detector the multiwire proportional chamber (MPC) was used. In 70th the MPC did not have a high factor of gas amplification and accordingly they were not highly sensitive to the low-energy radiation. Due to the development of technology of multiwire detectors there both the factors of gas amplification and the sensitivity were increased. New types of detectors were developed as well, these were, first of all, the various avalanche detectors having high factors of internal amplification of the order of $10^{6-7}$ /3-11/. Avalanche detectors having a single-photonic and a single-electronic sensitivity are capable to register a separate electron with practically zero energy. At such high level of sensitivity the characteristic of converters become determining.

In our earlier calculations, we took into account only electrons with energies higher than 29 keV /12-13/, and only electrons of internal conversion and Auger electrons, radiated from K-shell, with the energy 34.9 keV were taken into account. The minimal run of electrons with the energies of 29 keV in gadolinium is 4.7 microns. Not only we but also other authors /14-15/ made the same choice of energies of electrons. Thus the low-energy Auger electrons, these are electrons irradiated from L-shell with the energy of 4.84 keV and electrons from M-shell with the energy of 0.97 keV, were not taken into account. These electrons have rather small runs in gadolinium; namely, 0.3 microns (4.84 keV) and 0.04 microns (0.97 keV). They bring the small contribution to the total efficiency at use of converters made from natural gadolinium as the length of free runs of neutrons in natural gadolinium makes tens microns. At the same time their contribution becomes essential at use of converters made from 157 isotope of gadolinium as the length of free runs of neutrons in this material does not exceed 2-3 microns and this length becomes comparable with length of run of electrons.

In the present paper the analysis of influence of nano-scale layers of converters to the total efficiency of registration of thermal neutrons, on the example of converters made from natural gadolinium and its 157 isotope, is carried out. Contributions of low-energy Auger electrons having the maximal run in gadolinium 300 and 40 nanometers, and average run 100 and 15 nanometers, to the total efficiency of neutron converters were estimated.

2. Physical bases and model calculations

Among the solid-state converters of thermal neutrons converters made on the basis of gadolinium and especially its 157 isotope have the highest section of interaction. At calculations of the efficiency of registration of thermal neutrons by gadolinium converters three the most essential factors are taken into account, these are the probability of capture of neutrons by the material of the converter (absorption), the probability of radiation of electrons of internal conversion and Auger electrons, and also the probability of an escape of the secondary electrons from the material of the converter in view of thickness of the converter and angle of emission. Product of these three probabilities gives the required efficiency.

At radiating capture of neutrons by nucleus of gadolinium the whole cascade of gamma-quanta with the total energy of 7937.33 keV is emitted. In total 390 lines with the energy range from 79.5 up to 7857.670 keV with the intensities of lines from $2 \cdot 10^{-3}$ up to 139 gamma-quanta on 100 capture neutrons /17/ is emitted. There are low-energy gamma-quanta in this spectrum, and during their emission electrons from the nuclear shell (electrons of internal conversion) are radiated with a high probability. The nucleus eliminates its excitation by irradiation of gamma-quantum, but there also a close located electron could be emitted. As usual the K-electron (an electron of K-shell) is emitted, but also electrons from the higher shells (L, M, N and so on) could be emitted as well /18/. Vacancy of an electron (an electronic hole), formed in this process, is filled by another electron from a higher level. This process is accompanied by radiation of X-ray quantum or radiation of Auger electron.

Energies of emitted electrons of internal conversion are determined by the energies of irradiated gamma-quanta and bond energies of electrons on nuclear shells. During the electron emission an electron hole is created, which is filled by electrons from a higher levels. The process of filling of the vacancies is accompanied by X-ray radiation; its energy is equal to the difference of bond energies at the appropriate levels. Energy of excitation of atom could be also eliminated due to emission of Auger electron. These electrons are emitted instead of X-ray quanta and have energies equal to the energy of X-ray quantum minus bond energy of an electron at the appropriate level.

There are no data in the literature and databases on the runs of electrons with energies less than 10 keV in various materials. All data begin with this energy /19/. For calculation of run of Auger electrons in gadolinium we had to extrapolate the available data up to practically zero energy. Results of extrapolation are given in the Fig. 1. One can see from the figure, that the maximal run of electrons with the energy 4.84 keV makes 0.3 microns, and for electrons with the energy 0.97 keV makes 0.04 microns. These runs are rather small, especially in comparison with the neutron runs in natural gadolinium. At the same time for the converters made from 157 isotope of gadolinium the run in 0.3 microns makes approximately 10 % from the length of free run of neutrons and it can increase the efficiency of converters essentially. These small runs in 300 and 40 nanometers dictate the necessity to choose the step of iteration no more than 10 nanometers.

There are different data in the literature on the quantity of Auger electrons, so the quantity of electrons from the K-shell varies from 10 up to 14 % /20,21/, these data are normalized to the intensity of X-ray radiation of Ka1 (KL3) line, which is the most intensive one and which intensity is chosen for 100 %. The data on the quantity of electrons from L-shell varies even more from 150 up to 200 /20, 21/. For the electrons from the M-shell there are no data in general. At a choice of the quantity of electrons from the M-shell we started with the assumption that the quantity of Auger electrons grows with the distance of shell from a nucleus. So the quantity of electrons grows approximately 15 times at the transition from K to L shell. This tendency is kept further, i.e. from L to M and further N and O-shells. Due to Auger effect external electronic shells are fleeced from electrons practically completely. During the radiating capture of neutrons by nucleus of gadolinium there the irradiation of electrons of internal conversion takes place. The probability of emission of electrons of internal conversion makes approximately 60 %. Auger electrons accompany the emission of electrons of internal conversion, since during the irradiation of electrons of internal conversion there the vacancies on the electronic shells are formed, which are filled by electrons from higher shells. This process is accompanied by X-ray radiation and Auger electrons. In our calculations, at normalizing to the quantity of neutrons, during the calculation of quantity of Auger electrons the quantity of electrons is necessary to normalize not to Ka1 (KL3) an X-ray line, but to the quantity of electrons of internal conversion, i.e., it depends on factor of internal conversion.

In the table 1 the most intensive lines of electrons emitted during radiating capture of thermal neutrons by nucleus of gadolinium are presented.

Tab. 1

| Energy of electrons (keV) | Output of electrons 1/100 n | Electron run in gadolinium (microns) | Energy of initial gamma-quantum | Comment, levels |
|---|---|---|---|---|

|  | (error) |  |  |  |
|---|---|---|---|---|
| 0.97 | >200 | 0,04 |  | M- Auger |
| 4.84 | 97(47) | 0.3 |  | L- Auger |
| 29.3 | 35.58 | 4.7 | 79.51 | K |
| 34.9 | 7.9(4) | 6,29 |  | K- Auger |
| 71.7 | 5.57 | 20.7 | 79.51 | L |
| 78 | 1,2 | 23.78 | 79.51 | M |
| 131.7 | 6.96 | 55.70 | 181.93 | K |
| 174.1 | 0,99 | 86.27 | 181.93 | L |
| 180.4 | 0.21 | 91.23 | 181.93 | M |
| 205.4 | 0.14 | 111.47 | 255.66 | K |
| 227.3 | 0.16 | 130.27 | 277.54 | K |
| 729.9 | 0.03 | 649.38 | 780.14 | K |
| 893.85 | 0.06 | 830.05 | 944.09 | K |
| 911.8 | 0.04 | 849.83 | 960 | K |
| 926.8 | 0.03 | 866.35 | 975.4 | K |

Converters made from 157 isotope of gadolinium have an abnormal high section of interaction to thermal neutrons so the section makes 253 778.40 barn for neutrons with the wavelength of 1.8 Å. The section strongly grows with the increase of wavelength of neutrons. Converters with the thickness of 2 microns absorb more than 80 % of neutrons with the wavelength of 1.8Å and more than 90 % of neutrons with the wavelengths more than 3Å, Fig. 2.

Another situation develops at use of converters made of natural gadolinium. So the section of interaction makes 48 149.41 barn for neutrons with the wavelength of 1.8 Å. Attenuation up to 80 % of neutron beam (1.8 Å) takes place at the thickness more than 12 microns.

The analysis of curves shows, that at use of 157 isotope of gadolinium, the small run of Auger electrons (< 0.3 microns) can increase the total efficiency of converters essentially.

We carried out calculations of the efficiency of gadolinium foils, at use of natural gadolinium and its 157 isotope, for four fixed neutron wavelengths neutrons depending on thickness of converters. In detail the procedure of calculation is given in papers /12,13/. The total efficiency of converters increases in view of the contribution of low-energy Auger electrons. For the converters made from 157 isotope of gadolinium the total efficiency is increased more than by 10 %. For the converters made from natural gadolinium the increase is not so appreciable. With increase of neutron wavelength the difference in the efficiencies is increased. Especially well the difference is visible at use of 157 isotope of gadolinium. It is shown, that at calculations of the efficiency of converters on the basis of 157 isotope of gadolinium it is necessary to take into account the nano-scale layers, the minimal step of iterations should be no more than 10 nanometers.

The received data were compared with the experimental data given in the paper /18/. In this work the experimental data on the efficiency of detecting of neutrons emitted in a back hemisphere for 6 different energies are received and their comparisons with calibrated $^3$He counter are conducted. In this work converters made from natural gadolinium and enriched up to 90.5 % 157Gd were used. Work /18/ was carried out on the reactors of Atominstitut in Vienna (ATI) and the ILL Grenoble.

One can see from the Fig. 3, that in the case of taking into account of the contribution of low-energy Auger electrons (electrons with the energies higher than 0.93 keV) the results of our calculations well coincide with the experimental data. This concurrence is well visible for the converters made from 157 isotope of gadolinium. If one do not to take into account the low-energy electrons, the curve lays much below the experimental data. Such good consent of theoretical calculations with the experimental data testifies to the correctness of models for calculations and theoretical preconditions.

3. Conclusion

Earlier in our calculations we took into account only electrons with the energies higher than 29 keV. Thus there the low-energy Auger electrons were not taken into account; these are Auger electrons from the L-shell with the energy 4.84 keV and Auger electrons of the M-shell with the energy 0.97 keV. These electrons have rather small runs in gadolinium; these are 0.3 microns (4.84 keV) and 0.04 microns (0.97 keV). They bring the small contribution to the total efficiency at use of converters made from natural gadolinium as the lengths of free runs of neutrons in natural gadolinium makes tens micron. At the same time their contribution becomes essential at the use of converters made from 157 isotope of gadolinium as the lengths of free runs of neutrons there does not exceed 2-3 microns and this length become comparable with lengths of runs of electrons.

Calculations of efficiency of registration of thermal neutrons by foils made of natural gadolinium and its 157 isotopes are carried out. In the calculations the low-energy electrons were taken into account. Taking into account of the contribution of low-energy Auger electrons results in growth of the total efficiency of converters. For converters made from 157 isotope of gadolinium the total efficiency is increased more than by 10 %. For converters made from natural gadolinium the increase is not so appreciable. With increase of neutron wavelength the difference in the efficiencies is increased.

Especially well the difference is visible at use 157 isotope of gadolinium. It is shown, that at calculations of efficiency of converters on the basis of 157 isotope of gadolinium it is necessary to take into account the nano-scaled layers, and the minimal step of iterations should be no more than 10 nanometers. The received results are well coordinated to experimental data.

Such good consent of theoretical calculations with the experimental data testifies to the correctness of the chosen models and theoretical preconditions.

This research was conducted under the support of the International Science and Technology Center, Project T-1157.

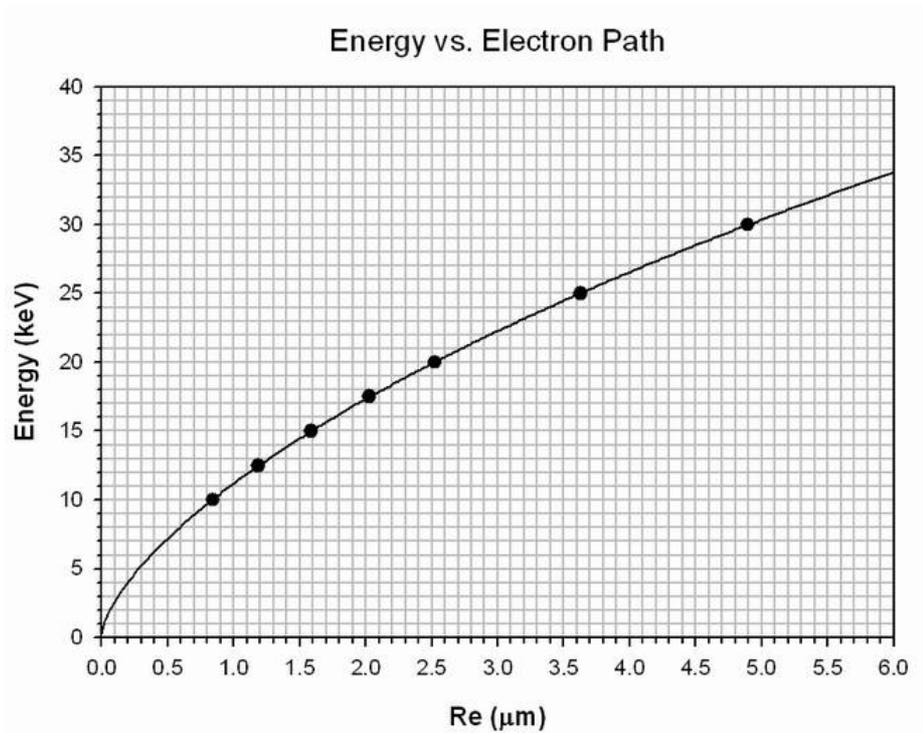

Fig. 1 Curve of dependence of run of electrons in gadolinium depending on their energy, for a low-energy range.

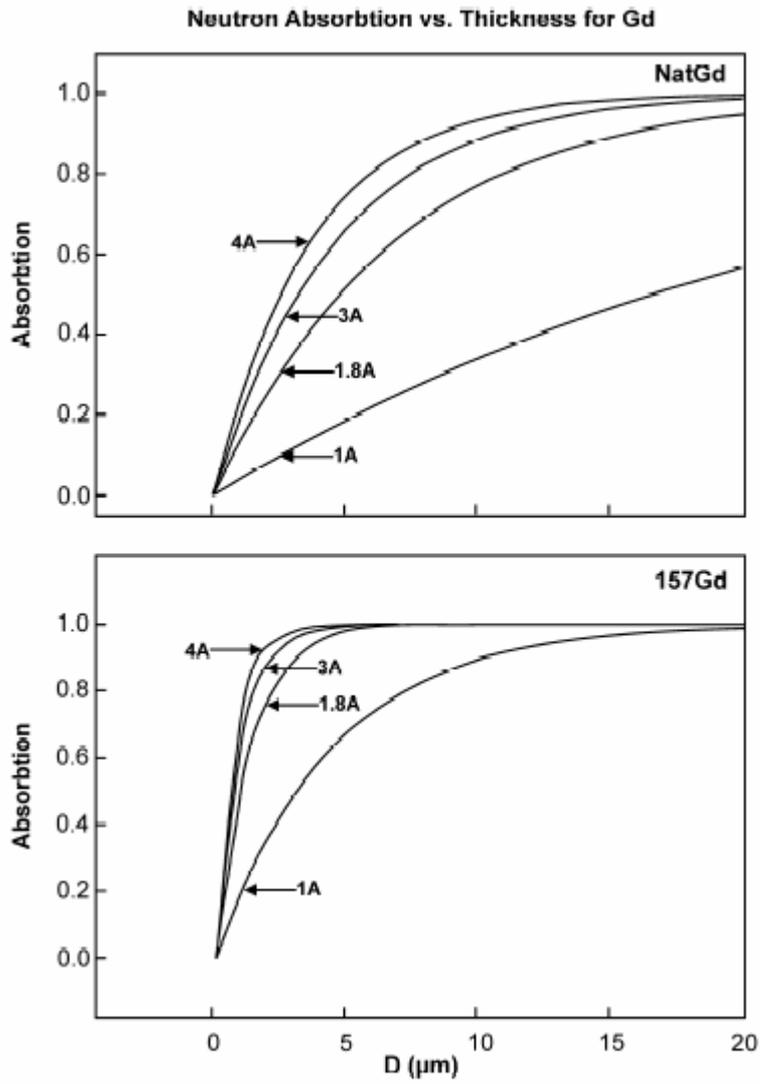

Fig. 2 Curves of absorption of neutrons for the wavelengths of 1, 1.8, 3 and 4 Å for natural gadolinium and its 157 isotope.

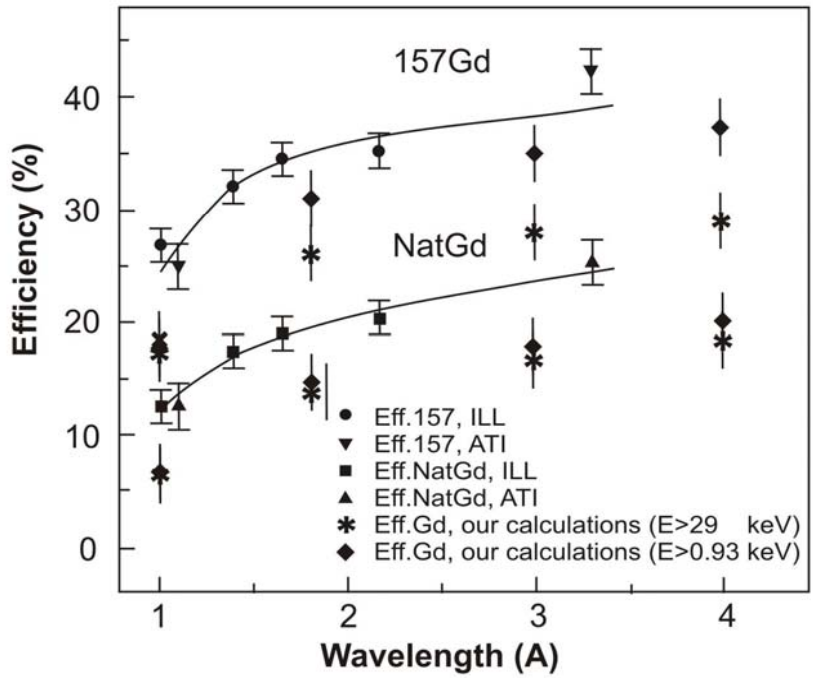

Fig. 3 Comparison of results of our calculations with the experimental data presented in the paper /18/. Calculations are carried out for two boundary values of the energies of taken into account electrons; i.e. for the values of energy higher than 29 keV and higher than 0.93 keV.